\documentclass[a4paper]{jpconf}
\usepackage{graphicx}

\begin{document}
\title{Are nonlinear discrete cellular automata compatible with quantum mechanics?}

\author{Hans-Thomas Elze}

\address{Dipartimento di Fisica ``Enrico Fermi'', Universit\`a di Pisa,  
Largo Pontecorvo 3, I-56127 Pisa, Italia}

\ead{elze@df.unipi.it} 

\begin{abstract}
We consider discrete and integer-valued cellular automata (CA). A particular class of which comprises ``Hamiltonian CA'' with equations of motion that bear  similarities to Hamilton's equations, while they present discrete updating rules. The dynamics is {\it linear},  
quite similar to unitary evolution described by the Schr\"odinger equation. 
This has been essential in our construction of an 
invertible {\it map} between such CA and continuous quantum mechanical models, which incorporate a fundamental {\it discreteness} scale. Based on   
Shannon's sampling theory, it leads, for example, to a one-to-one relation between  
quantum mechanical and CA conservation laws. The important issue of linearity of the 
theory is examined here by incorporating higher-order nonlinearities into the 
underlying action. These produce inconsistent nonlocal (in time) effects when trying to  describe continuously such nonlinear CA. Therefore, in the present framework, only linear 
CA and local quantum mechanical  dynamics are compatible.  
\end{abstract}

%%%%%%%%%%%%
%%%%%%%%%%%%
\section{Introduction}
A novel analysis of quantum mechanics (QM) --- which aims at  
redesigning the foundations of quantum theory  --- has recently been proposed  
by G.  't\,Hooft \cite{tHooft2014}. 
The hope for a comprehensive theory expressed in this far-reaching  
considerations is founded on the observation that quantum 
mechanical features arise in a large variety of deterministic ``mechanical'' 
models.  
While practically all of these models have been singular cases, i.e., which cannot easily be 
generalized to cover a realistic range of phenomena incorporating interactions, 
CA promise to provide the necessary versatility \cite{PRA2014,EmQM13,MyReview}. 
For an incomplete list of various earlier attempts 
in this field, see, for example,  
Refs.\,\cite{H1,H2,H3,Kleinert,Elze,Groessing,Jizba,Mairi,Isidro} and further 
references therein. 

The linearity of quantum mechanics (QM) is a fundamental feature, which is   
particularly visible in the Schr\"odinger equation (leaving aside here models 
which attempt to describe measurement processes dynamically). 
This is independent of the 
particular object  under study, provided it is sufficiently isolated from 
anything else, and is naturally reflected in the superposition principle. 
Thus, linearity  entails the ``quantum essentials''  interference and entanglement.    

The linearity of QM has been questioned and nonlinear modifications 
have been proposed earlier --- not only as suitable approximations for complicated 
many-body dynamics, but especially in order to test experimentally the  
robustness of QM against such {\it nonlinear deformations}. 
This has been thoroughly discussed by T.F. Jordan who presented a proof 
`from within' quantum theory that the theory has to be linear, given the 
{\it separability}   
assumption ``... that the system we are considering can be described as
part of a larger system without interaction with the rest of the larger 
system.''\cite{Jordan}

In our recent work, we have considered a {\it discrete  
dynamical theory}, which deviates notably from quantum theory, 
at first sight. 
However, we have shown with the help of sampling theory 
that the deterministic mechanics of the class of 
Hamiltonian CA can be 
related to QM in the presence of a fundamental time scale. This 
relation appears to demonstrate that consistency of the action principle 
of the underlying discrete dynamics implies, in particular,  the linearity of both theories. 

We will review some pertinent results, followed by  a more detailed 
study of the undesirable consequences that we are confronted with, 
if we generalize the 
action principle by incorporating {\it genuine nonlinearities} and  
try to enlarge the proposed class of CA in this way. 
\vskip 0.2cm 

The CA approach may offer additional insight into  interference and entanglement, in the 
limit where the discreteness scale can be treated as sufficiently small.  
Furthermore, future developments are conceivable which can address the {\it dynamics} 
of QM measurement processes, about which standard quantum 
theory remains silent. 

%%%%%%%%%%%%%%%%%%%%%%%%%%
\section{An action principle for cellular automata} 
In the following, we recall briefly the dynamics of a class of discrete CA 
automata introduced earlier, 
which bears remarkable similarity with Hamiltonian dynamics in the continuum 
on one side and with QM on the other \cite{PRA2014,EmQM13,MyReview}. 

We describe the state of a classical CA with countably many degrees 
of freedom by discrete {\it integer-valued} ``coordinates''   
$x_n^\alpha ,\tau_n$ and ``conjugated momenta'' $p_n^\alpha ,\pi_n$, where 
$\alpha\in {\mathbf N_0}$ denote different degrees of freedom and $n\in {\mathbf Z}$  
different states.  
The {$x_n$ and $p_n$ might be higher-dimensional vectors, while $\tau_n$ and 
${\cal P}_n$ are assumed one-dimensional. --- The ``coordinate'' $\tau_n$ has been 
separated from the $x_n^\alpha$'s (correspondingly $\pi_n$ from the $p_n^\alpha$'s), 
since this degree of freedom allows to represent a {\it dynamical time} variable, 
discussed in \cite{PRA2014,EmQM13,Lee,TimeMach}, with further references given there. 
   
For any one of the dynamical variables, say $f_n$, the finite difference operator 
$\Delta$ is defined by: 
\begin{equation}\label{findiff}
\Delta f_n:=f_n-f_{n-1} 
\;\;. \end{equation} 
Furthermore, we introduce the quantities (assuming the summation convention for 
Greek indices, 
$r^\alpha s^\alpha\equiv\sum_\alpha r^\alpha s^\alpha$): 
\begin{eqnarray}\label{an} 
a_n&:=&c_n\pi_n 
\;\;, \\ \label{Hn} 
H_n&:=&\frac{1}{2}S_{\alpha\beta}(p_n^\alpha p_n^\beta+x_n^\alpha x_n^\beta  )
+A_{\alpha\beta}p_n^\alpha x_n^\beta + R_n
\;\;, \\ \label{An} 
{\cal A}_n&:=&\Delta \tau_n (H_n+H_{n-1})+a_n
\;\;, \end{eqnarray} 
where the constants, $c_n$, the symmetric, $\hat S\equiv\{ S_{\alpha\beta}\}$,  and the 
antisymmetric, $\hat A\equiv\{ A_{\alpha\beta}\}$, matrices are all integer-valued. 
The definition (\ref{an}) determines the 
behaviour of the variable $\tau_n$ and only the most simple 
choice (involving a single constant) will be relevant for our purposes, cf. below. 
$R_n$ stands for higher than second powers in $x_n^\alpha$ or  $p_n^\alpha$.  --- 
Discussion of such genuine {\it nonlinearities} is the aim of the present  
note and will be resumed in due course. 
 
Based on these definitions, we introduce the {\it integer-valued} CA {\it action} by:  
\begin{equation}\label{action} 
{\cal S}:=
\sum_n[(p_n^\alpha +p_{n-1}^\alpha )\Delta x_n^\alpha 
+(\pi_n+\pi_{n-1})\Delta\tau_n
-{\cal A}_n]  
\;\;. \end{equation} 
Furthermore, we consider      
{\it integer-valued variations} $\delta f_n$ to be applied to a polynomial $g$: 
\begin{equation}\label{variation} 
\delta_{f_n}g(f_n):=[g(f_n+\delta f_n)-g(f_n-\delta f_n)]/2\delta f_n 
\;\; , \end{equation} 
and $\delta_{f_n}g\equiv 0$, if $\delta f_n=0$. ---  In terms of these notions, 
we postulate the following variational principle for the CA dynamics.     
\vskip 0.25cm \noindent 
\underline{\it The CA Action Principle}: \hskip 0.1cm   
The discrete evolution of a CA is determined by the stationarity of its 
action under arbitrary integer-valued variations of all 
dynamical variables, $\delta {\cal S}=0$.\hfill $\bullet$ \vskip 0.25cm  

Some characteristic features of this {\it CA Action Principle} deserve to be mentioned:  
\vskip 0.15cm \noindent  
{\bf i)} Variations  of terms that are 
{\it constant, linear, or quadratic in integer-valued variables}  yield 
analogous results as  
standard infinitesimal variations of corresponding  terms in the continuum. 
\vskip 0.15cm \noindent 
{\bf ii)} While infinitesimal variations do not conform with integer valuedness, 
there is {\it a priori} no restriction of integer variations, hence {\it arbitrary 
integer-valued variations} must be admitted. 
\vskip 0.15cm \noindent 
{\bf iii)} However, for arbitrary  variations 
$\delta f_n$, the {\it remainder of higher powers}, $R_n$ in $H_n$, which 
enters the action, has to vanish for consistency.  
Otherwise the number of equations of motion 
generated by variation of the action, according to Eq.(\ref{variation}), 
would exceed the number of variables. (A suitably chosen $R_0$ or a 
sufficient small number of such remainder terms can serve to encode  
the {\it initial conditions} for CA dynamics.)   \vskip 0.1cm  

As we observed in earlier work, these features seem to be essential when  
constructing a map between the considered CA and equivalent quantum mechanical 
models based on sampling theory, {\it cf.} Sec.\,3.  We will try to illuminate this 
in Sec.\,4., by studying  a generalization of the variational principle 
incorporating nonlinearities and by pointing out consequences of such nonlinearities 
which obstruct this way to arrive at QM models from CA.  

%%%%%%%%%%%%%%%%%%%%%%%%%%%%%%%%%%
\subsection{The CA equations of motion and conservation laws} 
We now apply the {\it CA Action Principle} to the action ${\cal S}$ (keeping $R_n\equiv 0$ 
for the moment), which yields the following discrete equations of motion:   
\begin{equation}\label{xdotCA} 
\dot x_n^\alpha\;=\;\dot\tau_n(S_{\alpha\beta}p_n^\beta +A_{\alpha\beta}
x_n^\beta ) 
\;\;,\;  
\;\;\dot p_n^\alpha \;=\;-\dot\tau_n(S_{\alpha\beta}x_n^\beta -A_{\alpha\beta}p_n^\beta ) 
\;\;, \end{equation}
\begin{equation}\label{taudotCA} 
\dot\tau_n\;=\;c_n 
\;\;,\;  
\;\;\dot\pi_n\;=\;\dot H_n 
\;\;, \end{equation} 
introducing the suggestive notation $\dot O_n:=O_{n+1}-O_{n-1}\;$.   
We emphasize that all terms herein are integer-valued. The fact that we 
arrive at {\it finite difference equations} reflects the discreteness of the 
{\it automaton time} $n$ and their appearance has motivated the name 
{\it Hamiltonian CA}.   

The equations of motion are {\it time reversal invariant}, since  
the state $n+1$ can be calculated from knowledge of the 
earlier states $n$ and $n-1$ and the state $n-1$ from the later ones $n+1$ and $n$. ---  
Note that the $\dot\tau_n$ present parameters for the evolving $x,p$-variables, as  
a consequence of Eqs.\,(\ref{taudotCA}). More generally, $\dot\tau$ can play 
the role of a dynamically coupled {\it lapse function} in Eqs.\,(\ref{xdotCA}).  

Introducing the self-adjoint matrix $\hat H:=\hat S+i\hat A$,  
the Eqs.\,(\ref{xdotCA}) can be combined into: 
\begin{equation}\label{discrS} 
 \dot x_n^\alpha +i\dot p_n^\alpha =-i\dot\tau_n H_{\alpha\beta}
(x_n^\beta +ip_n^\beta ) 
\;\;, \end{equation} 
and its adjoint. Thus, we find here a {\it discrete analogue of Schr\"odinger's equation}, 
with $\psi_n^\alpha :=x_n^\alpha +ip_n^\alpha$ as amplitude of the 
``$\alpha$-component''  
of ``state vector'' $|\psi\rangle$ at ``time'' $n$ and with  $\hat H$ as  
{\it Hamilton operator}. (We will use QM terminology freely here and in the 
following.)  

Furthermore, there are {\it conservation laws} respected by the discrete equations of 
motion, or by Eq.\,(\ref{discrS}), which are in {\it one-to-one correspondence} with those 
of the corresponding Schr\"odinger equation in the continuum 
\cite{PRA2014,EmQM13,MyReview}. ---  
In particular, the Eqs.\,(\ref{xdotCA}) imply the following theorem.   
\vskip 0.25cm \noindent 
\underline{\it Theorem A}: \hskip 0.1cm For any matrix $\hat G$ that commutes with 
$\hat H$, $[\hat G,\hat H]=0$, there 
is a {\it discrete conservation law}: 
\begin{equation}\label{Gconserv} 
 \psi_n^{\ast\alpha}G_{\alpha\beta}\dot\psi_n^\beta +
\dot\psi_n^{\ast\alpha}G_{\alpha\beta}\psi_n^\beta =0 
\;\;. \end{equation}  
For self-adjoint $\hat G$,  with complex integer elements, 
this relation concerns real integers.\hfill $\bullet$  
\vskip 0.25cm \noindent 
\underline{\it Corollary A}: \hskip 0.1cm For $\hat G:=\hat 1$, the Eq.\,(\ref{Gconserv}) 
implies a {\it conserved constraint} on the state variables:  
\begin{equation}\label{psiconserv} 
 \psi_n^{\ast\alpha}\dot\psi_n^\alpha +
\dot\psi_n^{\ast\alpha}\psi_n^\alpha =0 
\;\;. \end{equation}  
For $\hat G:=\hat H$, an {\it energy conservation} law follows.\hfill $\bullet$  
\vskip 0.25cm 

Note that Eqs.\,(\ref{Gconserv}) and (\ref{psiconserv}) {\it cannot} be  
``integrated'' as usual, since the {\it Leibniz rule} is modified here. 
Recalling $\dot O_n:=O_{n+1}-O_{n-1}$, we find, for example, 
$O_{n+1}O'_{n+1}-O_{n-1}O'_{n-1}=\frac{1}{2}(\dot  O_n[O'_{n+1}+O'_{n-1}]
+[O_{n+1}+O_{n-1}]\dot O'_n)$, instead of the product rule of differentiation. 

Furthermore, the continuum limit of the equations of motion and 
their conservation laws does not simply follow from letting  
the discreteness scale $l\rightarrow 0$. The integer valuedness 
of all quantities conflicts with continuous time and related derivatives. 
Hence, we need a more elaborate mapping, in order to relate CA to  
continuum models.  
\vskip 0.2cm 

In the following Sec.\,3., it will be shown that such an {\it invertible map} 
between the descriptions of discrete time Hamiltonian CA and of 
quantum mechanical objects evolving in continuous time can indeed be 
constructed, taking into account a fundamental discreteness scale $l$.

%%%%%%%%%%%%%%%%%%%%%%%%%%%%%%%%%%%%%
\section{The CA $\leftrightarrow$ QM map based on sampling theory}   
Despite the notable similarities of the Hamiltonian CA with QM systems, 
we may wonder whether the discreteness of a CA can be reconciled with 
a continuum description at all and, in particular, with QM?  

We have suggested earlier that especially wave functions, like other fields, could be 
{\it simultaneously discrete and continuous}, represented by sufficiently smooth  functions containing a finite density of degrees of freedom \cite{PRA2014,EmQM13}. 
Related ideas have been discussed  by T.D. Lee and collaborators 
and recently by A. Kempf in attempts to introduce a covariant ultraviolet cut-off 
into quantum field theories and, last not least, for gravity, see, for example, 
Refs.\,\cite{Lee,Kempf} with further references there (and in 
\cite{PRA2014,EmQM13}).   
However, {\it integer-valued CA} have first been connected with {\it structure of QM} 
from this perspective in our work.  

Information can have simultaneously
continuous and discrete character as pointed out by C.E. Shannon 
in his pioneering work \cite{Shannon}. This is routinely applied in signal 
processing, converting analog to  
digital encoding and {\it vice versa}. Sampling theory demonstrates  
that a bandlimited signal can be perfectly reconstructed, 
provided discrete samples of it are taken at the rate of at least twice the band 
limit (Nyquist rate) -- for an extensive review, see Ref.\,\cite{Jerri}. 

We shall make use of the basic version \cite{Jerri} of the \underline{\it Sampling Theorem}:  
\\ \noindent 
Consider square integrable {\it bandlimited functions} $f$, {\it i.e.}, which can be 
represented as $f(t)=(2\pi )^{-1}\int_{-\omega_{max}}^{\omega_{max}}\mbox{d}\omega\; 
\mbox{e}^{-i\omega t}\tilde f(\omega )$, with bandwidth $\omega_ {max}$. Given 
the set of amplitudes $\{ f(t_n)\}$ for the set  $\{ t_n\}$ of equidistantly spaced times  
(spacing $\pi /\omega_{max}$), the function $f$ is obtained for all $t$ 
by: 
\begin{equation}\label{samplingtheorem} 
f(t)=\sum_n f(t_n)\frac{\sin [\omega_{max}(t-t_n)]}{\omega_{max}(t-t_n)} 
\;\;. \end{equation} 

Since the CA state is labelled by the integer $n$, the {\it automaton time}, 
the corresponding discrete {\it physical time}  
is obtained by  multiplying with the fundamental scale $l$,  $t_n\equiv nl$, and the 
bandwidth by $\omega_{max}=\pi /l$. 

When attempting to {\it map invertibly} Eqs.\,(\ref{xdotCA}) on  
continuum equations, according to Eq.\,(\ref{samplingtheorem}), the nonlinearity 
on the right-hand sides is problematic, since the product of two 
functions, with bandwidth $\omega_ {max}$ each, is not a function with the same 
bandwidth. Therefore, we assume here that $\dot\tau_n$ is a 
constant and postpone the discussion of generic nonlinearities to Sec.\,4.  

Recalling the discrete time equation (\ref{discrS}), we insert  
$\psi_n^\alpha :=x_n^\alpha +ip_n^\alpha$, as before, and      
apply the {\it Sampling Theorem} to obtain the equivalent 
{\it continuous time equation}: 
\begin{equation}\label{modS}
(\hat D_l-\hat D_{-l})\psi^\alpha (t)=2\sinh (l\partial_t)\psi^\alpha (t)
=\frac{1}{i}H_{\alpha\beta}\psi^\beta (t) 
\;, \end{equation} 
employing the translation operator defined by $\hat D_Tf(t):=f(t+T)$ and 
implementing the natural choice $\dot\tau_n\equiv 1$, for all $n$. 
 
It appears that we recover the {\it Schr\"odinger equation}. However, it is modified 
in important ways, reflecting the presence of the scale $l$.  

First of all, by construction, the continuous time wave function  $\psi^\alpha$ is 
bandlimited (by $\omega_{max}$). Therefore, knowing 
the wave function at the discrete times of a set $\{ t_0+nl|n\in\mathbf{Z}\}$, with 
$t_0$ arbitrary, it can be reconstructed for all times  by a slight generalization of 
Eq.\,(\ref{samplingtheorem}). Furthermore, since Eq.(\ref{modS}) is of the form  
$f(t+l)=f(t-l)-i\hat Hf(t)$, it is sufficient to know $f$ at two times, say $t_0$ and 
$t_0-l$, in order to obtain it for all times of the set $\{ t_0+nl|n\in\mathbf{Z}\}$. 
Thus, we learn that {\it two initial conditions} (separated by a time step $l$) have to 
be specified to define the solution of Eq.\,(\ref{modS}). This agrees precisely with 
the requirements of the discrete description of the CA, {\it cf.} Sec.\,2.1. --- 
In order that sampling reproduces  
the {\it integer-valuedness} of the CA, both initial values have to be integer-valued. 

If instead the modified Schr\"odinger equation is written in terms of the 
infinite series of odd powers of time derivatives, it might give the false impression that 
an infinity of initial conditions are required. In any case, the higher-order derivatives are 
negligible for low-energy wave functions, which vary little with respect to the cut-off 
scale, {\it i.e.} $|\partial^k\psi /\partial t^k|\ll l^{-k}=(\omega_{max}/\pi )^k$. 

Secondly, the bandlimit $|\omega |\leq\omega_{max}$ leads to an 
{\it ultraviolet cut-off} of the energy $E$ of stationary states of the generic form $\psi_E(t):=\exp (-iEt)\tilde\psi$. Diagonalizing the self-adjoint  Hamiltonian,  
yields an eigenvalue equation and a {\it modified dispersion relation}: 
\begin{equation}\label{dispersionrel}
E_\alpha =l^{-1}\arcsin (\epsilon_\alpha /2)=
(2l)^{-1}\epsilon_\alpha [1+\epsilon_\alpha^{\;2}/24+\mbox{O}(\epsilon_\alpha^{\;4})]
\;\;, \end{equation}   
where $\{\epsilon_\alpha\}$ is the set of eigenvalues of the Hamiltonian; 
{\it e.g.}, $\alpha$ could label the momentum modes of a given  
spatial lattice.  We find that the spectrum $\{ E_\alpha\}$ is cut off by 
$|E_\alpha |\leq \pi /2l=\omega_{max}/2$, {\it i.e.} half the bandlimit. 
We have discussed further aspects of this result elsewhere \cite{MyReview,MyWigner14}.  
\vskip 0.2cm

Finally, the discrete CA conservation laws, {\it Theorem A} and {\it Corollary A} in Sec.\,2.1., 
Eqs.\,(\ref{Gconserv}) and (\ref{psiconserv}), respectively,  naturally have a counterpart 
in the continuum 
description obtained with the help of sampling theory. The absence of 
an ordinary time derivative in the discrete CA model, where only finite differences 
can play a role, leads to similar combinations of translation operators as on the 
left-hand side of the modified Schr\"odinger equation, Eq.\,(\ref{modS}), in the 
continuous time conservation laws. See  
Refs.\,\cite{EmQM13,MyReview}, where also related symmetries have been addressed.  

%%%%%%%%%%%%%%%%%%%%%%%%%%%%%%%%%%%
\section{The options for nonlinear Hamiltonian cellular automata} 
The general properties and certain quantum features, in particular, of the Hamiltonian CA that we recalled in the previous sections derive from the {\it CA Action Principle} 
introduced in Sec.\,2. A most important aspect that we observed has been the {\it linearity} 
of the equations of motion, Eqs.\,(\ref{xdotCA}) or Eq.\,(\ref{discrS}). 

We have repeatedly pointed out that additional higher-order terms in the action,  
which would lead to 
nonlinear (in $x^\alpha_n,p^\alpha_n$ or $\psi^\alpha_n$) terms in the equations of 
motion, would simultaneously enlarge the set of equations. 
This differs markedly from what one is used to in applications of variational principles 
based on the continuum, where ordinary differential calculus is available.  
Presently, due to the absence of infinitesimals and ensuing necessity to admit 
arbitrary variations, the CA dynamics tends to become overdetermined. --- 
As we shall see in the following, however, a closer look at this problem reveals some interesting information on the availability and consequences of nonlinear extensions.   

%%%%%%%%%%%%%%%%%%%%%%%%%%%
\subsection{Generalizing the variational derivative}
The additonal equations of motion, which can render the equations of motion 
inconsistent since overdetermined, are related to two aspects 
of the dynamics. One comprises the additional higher than quadratic 
powers of dynamical variables in the action, which kind of terms we summarized 
by $R_n$ in definition (\ref{Hn}). While the other consists in the 
{\it arbitrary integer-valued variations} $\delta f_n$ of all dynamical variables 
present in the action, which are allowed by the variational principle. 
They are applied according to the definition of Eq.\,(\ref{variation}) and this  
can produce additional terms which involve powers of $\delta f_n$. The coefficients 
of such terms all have to vanish independently, leading to additonal 
equations of motion. The resulting enlarged set of equations, as compared to a given  
number of variables, will be overdetermined in generic cases. 

This problem can be avoided by suitably generalizing the definition 
of the variations. Replacing Eq.\,(\ref{variation}), we define here: 
\begin{equation}\label{genvariation} 
\delta_fg^{(N)}(f):=\sum_{k\geq 1}\gamma_k[g^{(N)}(f+m_k\delta f)-g^{(N)}(f-m_k\delta f)]
/2\delta f  
\;\;, \end{equation} 
where $f$ stands for a dynamical variable entering the $N$-th degree polynomial 
$g^{(N)}$ and  
$\gamma_k$ and $m_k$ ($m_k\neq m_{k'}$, for $k\neq k'$)  are constant real and positive integer-valued coefficients, respectively, to be determined. --- 
Namely, we aim to arrange these coefficients in such a way that 
$\delta_fg^{(N)}(f)=\bar g^{(N-1)}(f)$; thus, the variation results in a 
polynomial $\bar g^{(N-1)}$ of degree $N-1$ and all other possible terms proportional 
to powers of  $\delta f$ {\it cancel by construction}.   

This eliminates the possibility of having an overdetermined set of 
(generally nonlinear) equations of motion.  --- 
In order to proceed, we write the polynomial $g^{(N)}$ explicitly, 
\begin{equation}\label{gN}  
g^{(N)}(f):=g_0+g_1f^1+\;\dots\; +g_Nf^N 
\;\;, \end{equation}  
and expand the difference appearing in Eq.\,(\ref{genvariation}): 
\begin{eqnarray}\label{diff} 
[g^{(N)}(f+m_k\delta f)&-&g^{(N)}(f-m_k\delta f)]/2\;=\; 
g_1\cdot m_k\delta f\;+\;g_2\cdot 2m_kf\delta f 
\nonumber \\ [1ex] 
&+&g_3\cdot\left (3m_kf^2\delta f+(m_k\delta f)^3\right ) 
\nonumber \\ [1ex] 
&+&g_4\cdot\left (4m_kf^3\delta f+4f(m_k\delta f)^3\right ) 
\nonumber \\ [1ex]
&+&g_5\cdot\left (5m_kf^4\delta f+10f^2(m_k\delta f)^3+(m_k\delta f)^5\right ) 
\nonumber \\ [1ex] 
&+&\;\dots 
\;\;. \end{eqnarray} 
Note that the terms $\propto\delta f$ correspond to the ones obtained by  
ordinary differentiation of the polynomial. However, since any one    
dynamical variable $f$ of our CA is integer-valued,  such derivatives are to be 
interpreted only as a symbolical notation in the following. 
Thus, we have $\delta_fg^{(N)}(f)=
\sum_k\gamma_km_k\cdot (\mbox{d}/\mbox{d}f)g^{(N)}(f)+\dots\;$, where
the additional terms  
involving powers of $\delta f$ are not spelled out. 
The point of our considerations is that the latter can be made to vanish 
always by suitably adjusting the  coefficients $\gamma_k$ and $m_k$,  if we restrict 
the maximal order of polynomials to be dealt with.   

For illustration, we consider {\it all} polynomials of order $\leq 4$, {\it i.e.} $g^{(4)}$.  
Here, the terms $\propto\delta f^3$ are eliminated 
always, {\it cf.} Eq.\,(\ref{diff}), if the following condition is fullfilled: 
\begin{equation}\label{g5} 
\sum_{k\geq 1}\gamma_k(m_k)^3\stackrel{!}{=}0 
\;\;, \end{equation} 
A solution is provided by: $m_1=1$, 
$m_2=m\geq 2$, $\gamma_1=1/(1-m^{-2})$, $\gamma_2=-m^{-3}/(1-m^{-2})$, and 
all other coefficients vanishing. Thus, we obtain: 
$\delta g^{(4)}(f)=
(\mbox{d}/\mbox{d}f)g^{(4)}(f)$, {\it cf.} Eq.\,(\ref{genvariation}). This solution is 
sufficient for our purposes but not unique. 

By this elementary reasoning, we have obtained a satisfactory variational derivative, 
which avoids the problem of arriving at an overdetermined set of equations of motion. 
Our approach can be generalized to polynomials of arbitrary finite order; limitation 
to finite order being warranted by integer-valuedness of the variables.   

Consequently, a suitably generalized variational derivative can be employed 
in the {\it CA Action Principle}, such that consistent finite difference 
{\it equations of motion incorporating nonlinear potentials} result, which  
maintain the structural similarity with classical Hamilton's equations that 
we have seen in Sec.\,2.1.  

%%%%%%%%%%%%%%%%%%%%%%%%%%% 
\subsection{Problems that arise with nonlinearity} 
The generalized variational derivative of Eq.\,(\ref{genvariation}) can be employed to 
define a {\it Poisson bracket}, similarly as we discussed elsewhere 
\cite{MyReview,MyWigner14} --- Since this variational derivative acts practically like an ordinary derivative, there is apparently no 
need to restrict the related {\it algebra of observables} in any way. 

Yet it does remain consistent to consider only (linear or) {\it quadratic forms} in the 
dynamical variables, recovering the previous results and analogous symmetry properties 
as in QM \cite{MyWigner14,Heslot85}.  

However, once higher order polynomials (in $x^\alpha_n,p^\alpha_n$ or $\psi^\alpha_n$) are admitted in the action and equations of motion, or as observables, it is {\it not 
consistent} to limit the set of relevant polynomials at any 
finite order. For example, the Poisson bracket of two polynomials of order 
$N$ and $N'$, respectively, may result in a polynomial of order $N+N'-2>N,N'$. 
This is problematic, since arbitrarily high powers of integer-valued variables, 
and linear combinations thereof, generated in this way, will eventually lead 
to {\it divergent quantities}!

Thus, we observe here a qualitatively profound `bifurcation' in the properties of 
Hamiltonian CA tied to the presence or absence of  nonlinearities in their 
equations of motion. The preceding remarks seem to imply that  an algebra 
of observables cannot even be defined for the nonlinear case in an analogous way as  
in classical mechanics.   
To emphasize this point, we note that the kind of discrete 
or continuous {\it conservation laws} (and traces of QM unitary {\it symmetry}), discussed 
in Secs.\,2.1. or 3., see also Ref.\,\cite{EmQM13,MyReview}, will generally be absent in 
nonlinear CA.    

Further problematic features can be expected, when we consider the 
behaviour of nonlinear terms under the map 
relating the discrete description of CA and its continuum counterpart, employing Shannon's 
{\it Sampling Theorem} as in Sec.\,3., to which we turn shortly.  

%%%%%%%%%%%%%%%%%%%%%%%%%%%%%%%%
\subsubsection{A summary of properties of sinus cardinalis}
We will make use of several results concerning the {\it sinus cardinalis} 
function, $\mbox{sinc}(x):=\sin(x)/x$, which is employed in the reconstruction formula, 
Eq.\,(\ref{samplingtheorem}), and will be needed in the following. 

Let us define $s_n(t):=\mbox{sinc}[\pi (t/l-n)]$, which leads to the Fourier transform:  
\begin{equation}\label{snF} 
\int_{-\infty}^\infty\mbox{d}t\;\mbox{e}^{-i\omega t}s_n(t)
=l\theta (\pi /l-\omega )\theta (\pi /l+\omega )\mbox{e}^{-i\omega ln}
\;\;, \end{equation} 
where $\theta$ denotes the Heaviside step function. 
Thus, the function $s_n$ is {\it bandlimited}.  
Furthermore, it presents a ``nascent'' Dirac delta function, which has the properties: 
\begin{eqnarray}\label{Dirac1} 
&\;&\;\;\;\;\;\;\;l^{-1}\int_{-\infty}^\infty\mbox{d}t\;s_n(t)\;=\;1 
\;\;, \\ [1ex] \label{Dirac2} 
&\;&\lim_{l\rightarrow 0}\;l^{-1}\int_{-\infty}^\infty\mbox{d}t\;s_n(t)f(t)\;=\;f(0)
\;\;. \end{eqnarray} 
These results can be recovered with the help of the Fourier transform of 
$s_n$, Eq.\,(\ref{snF}), assuming that $f$ has  a well-behaved Fourier transform. 
Employing the inverse Fourier transformation of Eq.\,(\ref{snF}), we obtain the orthogonality relation:  
\begin{equation}\label{orthogonality} 
l^{-1}\int_{-\infty}^\infty\mbox{d}t\; s_m(t)s_n(t)=\delta_{mn} 
\;\;. \end{equation}  
Finally, as an example, we perform a summation by applying 
the {\it Sampling Theorem}, Eq.\,(\ref{samplingtheorem}): 
\begin{equation}\label{sincsum} 
\sum_{m\in\mathbf{Z}} s_n(ml-t')s_m(t)=s_n(t-t') 
\;\;, \end{equation} 
noting that the factor of $s_n$ under the sum can be interpreted as the 
function on the right-hand side sampled at the times $ml$; all functions here 
have identical bandwidth, {\it i.e.} $\omega_{max}=\pi /l$. 

%%%%%%%%%%%%%%%%%%%%%%%%%%%%
\subsubsection{Nonlinearity and nonlocality (in time)}
We now are ready to elaborate consequences of nonlinearity with respect to 
mapping the discrete equations of motion to the continuum 
desription thereof {\it via} application of Shannon's {\it Sampling Theorem}.  --- 
Suppose that the discrete analogue of the Schr\"odinger equation, 
Eq.\,(\ref{discrS}), includes a genuinely {\it nonlinear term}, {\it e.g.}:  
\begin{equation}\label{nonlinS} 
\dot\psi_n^\alpha=-iH_{\alpha\beta}\psi_n^\beta 
+M_{\alpha\beta\gamma}(\psi_n^{*\beta}+\psi_n^\beta)
(\psi_n^{*\gamma}+\psi_n^\gamma )
\;\;, \end{equation}
keeping $\dot\tau_n\equiv 1$ and where the coefficients $M_{\alpha\beta\gamma}$ are 
real and totally symmetric in the indices. 

A corresponding potential can be incorporated 
into the action and the additional nonlinearity in Eq.\,(\ref{nonlinS}) follows by applying 
a suitably generalized variational derivative, as discussed in Sec.\,4.1. 
We are not interested in the physical (ir)relevance of the present example of a cubic potential, but would like to see what happens with the nonlinear terms, {\it e.g.}  
$M_{\alpha\beta\gamma}\psi_n^\beta\psi_n^\gamma$,  
when the {\it Sampling Theorem} is applied to Eq.\,(\ref{nonlinS}), similarly as 
before in Sec.\,3.  

Suppressing irrelevant greek indices, we introduce  
$\psi_n=:\psi (t_n)$ and $\psi_n\psi_n=:\psi_{(2)}(t_n)$.  Through the reconstruction 
formula (\ref{samplingtheorem}) the {\it discrete time} values $\psi (t_n)$ and 
$\psi_{(2)}(t_n)$ are replaced by {\it continuous time} functions $\psi (t)$ and 
$\psi_{(2)}(t)$, respectively, and we would like to make the relation between the 
latter functions explicit.  
  
First of all, employing the orthogonality relation (\ref{orthogonality}), we  
invert the reconstruction formula: 
\begin{equation}\label{psin} 
\psi (t_n)=l^{-1}\int_{-\infty}^\infty\mbox{d}t\; s_n(t)\psi (t) 
\;\;. \end{equation} 
Which gives us simply: 
\begin{equation}\label{psi2n}  
\psi_{(2)}(t_n)=l^{-2}\int_{-\infty}^\infty\mbox{d}t'\; s_n(t')\psi (t')
\int_{-\infty}^\infty\mbox{d}t''\; s_n(t'')\psi (t'')  
\;\;. \end{equation} 
Applying now the reconstruction formula to $\psi_{(2)}(t_n)$, we obtain indeed a  
nonlinear relation between $\psi_{(2)}(t)$ and $\psi (t)$: 
\begin{equation}\label{psi2psi}
\psi_{(2)}(t)=l^{-2}
\int_{-\infty}^\infty\int_{-\infty}^\infty\mbox{d}t'\mbox{d}t''
\sum_{n\in\mathbf{Z}}s_n(t)s_n(t')s_n(t'')\psi (t')\psi (t'')  
\;\;, \end{equation} 
where we interchanged summation and integrations. Similarly  
as in Eq.\,(\ref{sincsum}), we can do the sum:  The function    
$s_{(2)n}(t',t''):=s_n(t')s_n(t'')=\mbox{sinc}[\pi (nl-t')/l]\mbox{sinc}[\pi (nl-t'')/l]$ is of the 
bandlimited kind and sampled here at the times $nl$; it is reconstructed by the summation
including the factor $s_n(t)$, in accordance with the {\it Sampling Theorem}. However, 
the bandwidths need to be considered carefully. By Fourier transformation, one
verifies that $s_{(2)n}$ has a doubled bandwidth,  
$\omega_{max}^{(2)}=2\pi/l=2\omega_{max}$, as compared to $s_n$, which one would 
guess. This is implemented by writing all appearances of $l$ in terms of $l/2$ and by  applying Eq.\,(\ref{samplingtheorem}) to yield the sum: 
\begin{eqnarray} 
\sum_{n\in\mathbf{Z}}s_n(t)s_{(2)n}(t',t'')
&=&\sum_{n\in\mathbf{Z}}
\mbox{sinc}[\pi \textstyle{\frac{nl/2-t/2}{l/2}}] 
\nonumber \\  
&\;&\;\;\;\cdot\;
\mbox{sinc}[\pi \textstyle{\frac{nl/2-t'/2}{l/2}}]
\mbox{sinc}[\pi \textstyle{\frac{nl/2-t''/2}{l/2}}]  
\nonumber \\ [1ex] \label{summation}
&=&\mbox{sinc}[\pi (t-t')/l]
\mbox{sinc}[\pi (t-t'')/l]  
\;\;. \end{eqnarray} 
Thus, we obtain: 
\begin{equation}\label{psi2psi}
\psi_{(2)}(t)=l^{-2}\Big (
\int_{-\infty}^\infty\mbox{d}t'\;
\mbox{sinc}[\pi (t-t')/l]\psi (t')\Big )^2 
\;\;, \end{equation} 
which expresses $\psi_{(2)}$ in terms of $\psi$. --- In the limit of vanishing discreteness 
scale, we obtain a simple quadratic term :
\begin{equation}\label{psi2psi1}
\lim_{l\rightarrow 0}\;\psi_{(2)}(t)=\big (\psi (t)\big )^2 
\;\;, \end{equation} 
with the help of Eq.\,(\ref{Dirac2}). This presents, of course, the expected result. 
It is {\it local in time}. 

However, we observe that the 
quadratic term on the right-hand side of Eq.\,(\ref{psi2psi}) consists in factors which are
{\it nonlocal in time}: the function $\psi$ is integrated over {\it all} times, 
weighted by the oscillating and slowly decaying {\it sinc} kernel. Inserting this result (and 
corresponding additional terms) into the continuous time version of the 
discrete analogue of a nonlinear Schr\"odinger equation, Eq.\,(\ref{nonlinS}), 
changes the character of this equation 
profoundly: it is {\it not anymore a consistent CA updating rule}! --- This should be 
contrasted with  
the left-hand side of Eq.\,(\ref{modS}), which is nonlocal as well. However, this 
nonlocality is rather mild  
and refers to two neighbouring instants in such a way that the linear equation can be solved 
forwards (or backwards) in time step by step, recalling the discussion of 
initial conditions following Eq.\,(\ref{modS}) in Sec.\,3. With the nonlocality 
here, due to an anharmonic potential, inserted on the right-hand side 
of the continuous time equation, this fails.   

One may also consider that 
application of the {\it Sampling Theorem} to a nonlinear finite difference equation, 
such as Eq.\,(\ref{nonlinS}), cannot lead to a consistent continuous 
dynamics, since the resulting linear 
and nonlinear terms have different bandwidths (unless an additional 
cut-off on nonlinear terms is introduced by hand). 
Which holds for any kind of polynomial nonlinearity.  

Thus, we expect nonlocality to be a  problem  
for {\it any} continuous description based on a form of sampling theory 
of an underlying discrete CA 
dynamics,  {\it unless the CA updating rules are linear} in the dynamical variables  
(as in Sec.\,2.). We anticipate this to be the case also if space is discrete, besides time, 
a situation which can be studied along 
similar lines \cite{David}.  
\vskip 0.2cm 

This leaves us with a speculative question: {\it Could it be that 
unitary linear evolution in continuous time --- which appears to hold universally in QM  
(leaving aside measurement processes) --- is dictated by a local perspective on more 
general, possibly nonlinear underlying CA dynamics?} In short: {\it Does locality filter 
for linearity?} 

%%%%%%%%%%%%
\section{Conclusions}
We  have briefly reviewed, in Sec.\,2., the description of a class of deterministic discrete 
cellular automata based on an action principle 
\cite{PRA2014,EmQM13,MyReview,MyWigner14}. In particular, we have recalled 
how this can be mapped with the help of Shannon's sampling theory 
\cite{Shannon,Jerri} on a continuous 
time picture, which resembles in many respects the description of nonrelativistic 
quantum mechanical objects. 

In Sec.\,3., we have pointed out  
the relation between the discrete CA updating rules, which are 
closely analogous to Hamilton's equations of motion in mechanics, and a modified 
Schr\"odinger equation, which includes additional terms due to the finite discreteness 
scale $l$ characterizing the CA (and leads to a modified dispersion relation, with 
energy bounded below and above). This extends 
to a one-to-one correspondence between the associated conservation laws, between 
continuous unitary symmetries and their discrete counterparts.    

Presently, in Sec.\,4., we have paid particular attention 
to  a generalization which incorporates polynomial nonlinearities into the action and 
equations of motion in a consistent way. 

This is motivated by earlier indications that only linear CA equations of motion could be  consistent \cite{PRA2014,EmQM13} --- which  appeared a surprising result in view of the 
essential linearity of QM \cite{Jordan}. It is a fact that too naive implementation of 
genuine nonlinearities tends to produce overdetermined CA updating rules. 

We have found here that introducing 
a consistent nonlinear generalization of the discrete CA dynamics leads to an 
inconsistent  
nonlocality (in time) of the corresponding continuous time description obtained with 
the help of sampling theory. Furthermore, we have argued that nonlinearity has a 
detrimental effect (through Poisson brackets based on discrete 
variational derivatives) on the algebra of observables, as far as it compares with QM 
\cite{Heslot85}. 
Both would severely spoil any attempt to construct models in the 
class of Hamiltonian CA which have QM features emerging at large scales 
({\it i.e.} discreteness scale $l\rightarrow 0$) from 
underlying CA dynamics.  

We conclude that 
nonlinearity must be excluded from the kind of CA model building 
studied here, in order to maintain locality in the continuous 
description to the extent possible, when ordinary derivatives 
are replaced by finite differences beneath.  

Yet one may wonder about   
effects of nonlinear CA processes, if they influence only some dynamical variables.   
Which could be different ones than those commonly described by a linear and local  evolution in quantum theory.  Is there room for stochastic phenomena that 
manifest themselves in QM measurements? 
  
In any case, in order to turn our observations of surprising connections between the 
properties of cellular automata and the quantum mechanical features of more 
familiar physical objects into a theory, as proposed by  G. 't\,Hooft \cite{tHooft2014}, 
several immediate problems call for attention. We should understand how composite 
systems fare in this context. 
Which is a prerequisite to analyze the CA analogue of QM measurement processes 
and, more generally, the role of superposition principle and entanglement. 
Above all, we need to understand how aspects of relativity and of the physics of 
spacetime come into play here. 

%%%%%%%%%%%%%%%%
\ack 
It is my pleasure to thank T.\,W.\,Kibble, J.\,J.\,Halliwell and C.\,M.\,Bender for discussions and N.\,Mavromatos and his collaborators for kindly inviting me to the very interesting 
conference DISCRETE 2014 (King's College, London, 2-6 December 2014).  

%%%%%%%%%%%%


\section*{References}
\begin{thebibliography}{99}

\bibitem{tHooft2014} 't~Hooft G 2014 The Cellular Automaton Interpretation of Quantum Mechanics. A View on the Quantum Nature of our Universe, Compulsory or Impossible?
{\it Preprint} arXiv:1405.1548 

\bibitem{PRA2014} Elze H-T 2014 Action principle for cellular automata and the linearity of 
quantum mechanics 
\emph{Phys. Rev.} A {\bf 89} 012111        

\bibitem{EmQM13} Elze H-T 2014 The linearity of quantum mechanics from the 
perspective of Hamiltonian cellular automata 
\emph{J. Phys.: Conf. Ser.} {\bf 504} 012004  

\bibitem{MyReview} Elze H-T 2015 Quantum Features of Natural Cellular Automata,  
in {\it Quantum physics and spacetime dynamics beyond peaceful coexistence} 
ed I Licata (London: Imperial College Press) in press  

\bibitem{H1} 't~Hooft G 1990 Quantization of discrete deterministic theories by 
Hilbert space extension
\emph{Nucl. Phys.} B {\bf 342} 471  

\bibitem{H2} 't~Hooft G, Isler K and Kalitzin S 1992 Quantum
field theoretic behavior of a deterministic cellular automaton 
\emph{Nucl. Phys.} B {\bf 386} 495  

\bibitem{H3} 't~Hooft G 1997 Quantum mechanical behaviour in a deterministic model  
\emph{Found. Phys. Lett.} {\bf 10} 105  

\bibitem{Kleinert} Haba Z and Kleinert H 2002  
Towards a simulation of quantum computers by classical systems  
\emph{Phys. Lett.} A {\bf 294} 139  

\bibitem{Elze} Elze H-T and Schipper O 2002  
Time without time: a stochastic clock model, 
\emph{Phys. Rev.} D {\bf 66} 044020  

\bibitem{Groessing} Gr\"ossing G 2004 From classical Hamiltonian flow to quantum theory:  derivation of the Schr\"odinger equation  
\emph{Found. Phys. Lett.} {\bf 17} 343   

\bibitem{Jizba} Blasone M, Jizba P, Scardigli F and Vitiello G 2009  
Dissipation and quantization for composite systems  
\emph{Phys. Lett.} A {\bf 373} 4106   

\bibitem{Mairi} Sakellariadou M, Stabile A and Vitiello G 2011  
Noncommutative spectral geometry, algebra doubling and the seeds of quantization  
\emph{Phys. Rev.} D {\bf 84} 045026  

\bibitem{Isidro} Acosta D, Fernandez de Cordoba P, Isidro J M and Santander J L G 2012 
An Entropic Picture of Emergent Quantum Mechanics  
\emph{Int. J. Geom. Meth. Mod. Phys.} {\bf 9} 1250048   

\bibitem{Jordan} Jordan T F 2006 Assumptions that imply quantum dynamics is linear, 
\emph{Phys. Rev.} A {\bf 73} 022101; 
{\it do.} 2009  Why quantum dynamics is linear  
\emph{J. Phys.: Conf. Ser.} {\bf 196} 012010    

\bibitem{Lee} Lee T D 1983 Can time be a discrete dynamical variable? 
\emph{Phys. Lett.} B {\bf 122} 217 

\bibitem{TimeMach} Elze H-T 2013 Discrete mechanics, time machines and hybrid systems  
\emph{EPJ Web of Conferences} {\bf 58} 01013  

\bibitem{Kempf} Kempf A 2010 Spacetime could be simultaneously continuous and discrete 
in the same way that information can  
\emph{New J. Phys.} {\bf 12} 115001 

\bibitem{Shannon} Shannon C E 1949 Communications in the presence of noise  
\emph{Proc. IRE} {\bf 37} 10  

\bibitem{Jerri} Jerri A J 1977 The Shannon Sampling Theorem --- Its Various  Extensions and 
Applications: A Tutorial Review  
\emph{Proc. IEEE} {\bf 65}  1565    

\bibitem{MyWigner14} Elze H-T 2014 
Quantumness of discrete Hamiltonian cellular automata
{\it EPJ Web of Conferences} {\bf 78} 02005 

\bibitem{Heslot85} Heslot A 1985 Quantum mechanics as a classical theory 
\emph{Phys. Rev.} D {\bf 31} 1341   

\bibitem{David} Gigli D 2014 Application of Shannon's Sampling Theorem in 
Quantum Mechanics {\it Master Thesis} (University of Pisa, December 2014) unpublished


%\bibitem{McKeeSmyth} J.F. McKee and C.J. Smyth, Integer symmetric matrices having all 
%their eigenvalues in the interval [-2,2], 
%\emph{J. Algebra} {\bf 317}, no. 1, 260  (2007) 

%\bibitem{tHooftOsc} G. 't Hooft, Quantum mechanics and determinism, 
%\emph{presented at PASCOS 2001}, arXiv:hep-th/0105105  

%\bibitem{ElzeOsc} H.-T. Elze, Emergent discrete time and quantization: relativistic particle 
%with extradimensions, 
%\emph{Phys. Lett.} A {\bf 310}, 110 (2003) 


\end{thebibliography}
\end{document}